\documentclass[twocolumn,superscriptaddress]{revtex4}

\usepackage{graphicx}% Include figure files
\usepackage{dcolumn}% Align table columns on decimal point
\usepackage{bm}% bold math
\begin{document}

\title{Macrospin model of incubation delay due to the field-like spin transfer torque}

\author{Samir Garzon}
\affiliation{Department of Physics and Astronomy and USC Nanocenter,
University of South Carolina, Columbia, SC 29208, USA.}
\author{Yaroslaw Bazaliy}
\affiliation{Department of Physics and Astronomy and USC Nanocenter,
University of South Carolina, Columbia, SC 29208, USA.}
\affiliation{Institute of Magnetism, Nat'l Acad. of Sciences, Kyiv,
Ukraine}
\author{Richard A. Webb}
\affiliation{Department of Physics and Astronomy and USC Nanocenter,
University of South Carolina, Columbia, SC 29208, USA.}
\author{Thomas M. Crawford}
\affiliation{Department of Physics and Astronomy and USC Nanocenter,
University of South Carolina, Columbia, SC 29208, USA.}
\author{Mark Covington}
\affiliation{Seagate Research, 1251 Waterfront Place, Pittsburgh, PA
15222, USA.}
\author{Shehzaad Kaka}
\affiliation{Seagate Research, 1251 Waterfront Place, Pittsburgh, PA
15222, USA.}

\begin{abstract}

We show that the absence of pre-switching oscillations (``incubation delay'') in magnetic tunnel junctions can be
explained within the macrospin model by a sizable field-like
component of the spin-transfer torque. It is further suggested that
measurements of the voltage dependence of tunnel junction
switching time in the presence of external easy axis magnetic fields
can be used to determine the magnitude and voltage dependence of the
field-like torque.

\end{abstract}

\date{\today}
\maketitle

A spin polarized electric current can transfer spin angular momentum to a magnetic material, generating a torque that can induce magnetization dynamics and even magnetization reversal~\cite{slonczewski_JMMM1996, berger_PRB1996}. While extensive measurements have tested the validity and limitations of the macrospin model with Slonczewski's spin transfer torque in metallic spin valves, recent experiments with magnetic tunnel junctions (MTJ's)~\cite{petit_PRL2007,kubota_NATPHYS2008,sankey_NATPHYS2008,li_PRL2008} have observed an additional ``field-like'' or ``perpendicular'' spin torque. The existence of a field-like torque was predicted for metallic spin valves ~\cite{heide_PRL2001,tserkovnyak_PRL2002,tserkovnyak_RMP2005}, but shown to be smaller than Slonkzewski's ``parallel'' torque~\cite{xia_PRB2002,zimmler_PRB2004}. For magnetic tunnel junctions, however, it was predicted that both torques could have similar magnitudes and that the field-like torque would have a quadratic dependence on voltage~\cite{theodonis_PRL2006,slonczewski_JMMM2007}. The observed field-like torques generally agree with theoretical predictions, but some controversies remain. For example, measurements in the frequency domain at low voltages~\cite{kubota_NATPHYS2008,sankey_NATPHYS2008} and measurements of switching currents at large voltages~\cite{li_PRL2008} report contradictory signs of the field-like term. This suggests that further theoretical analysis and experimental investigation are necessary to fully understand the origin and the functional form of this torque.

Here we report that the field-like torque can explain the absence of the pre-switching oscillations (``incubation delay'') found by Devolder {\it et al.}~\cite{devolder_PRL2008} in MTJ's. This observation could not be described within a macrospin model of magnetization reversal based on Slonczewski's spin transfer torque alone. Such a model predicts that pumping of the ferromagnetic resonance mode produces increasing oscillations in the resistance before switching~\cite{sun_PRB2000,krivorotov_SCIENCE2005}. However, by including the effects of a field-like spin torque term within a macrospin model we are able to reproduce the main features of the observed magnetization reversal: (i) a slow regular change of the resistance without oscillations preceding the switching, (ii) decaying oscillations of the resistance after switching, and (iii) similarity between magnetization reversal curves shifted so as to align their switching times (in our case the switching times $t_S$ are distributed between 0 and 10 ns). In addition, we propose time-domain experiments which could be used to measure the magnitude of the field-like spin torque term and its voltage dependence, settling the mentioned sign issues and motivating further theoretical investigation.

\begin{figure}[bt]
\begin{center}
\includegraphics[width=3.5in]{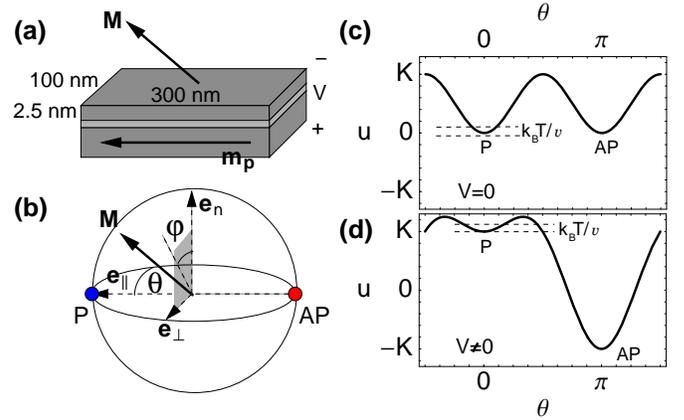}
\caption{(a) Schematic of rectangular MTJ showing the free layer magnetization $\mathbf{M}$ at an arbitrary orientation, and the reference layer magnetization $\mathbf{m_P}$ pinned along the easy axis. The polarity of the voltage V applied across the sample is shown. (b) Unit sphere describing the possible orientations of $\mathbf{M}$. Unit vectors normal to the sample ($\mathbf{e_n}$), and along the easy ($\mathbf{e_{\parallel}}$) and hard axis ($\mathbf{e_{\perp}}$) are shown. The orientation of $\mathbf{M}$ is described by $\theta$, the angle between the free and pinned magnetizations, and $\varphi$, the angle between $\mathbf{e_n}$ and the projection of $\mathbf{M}$ onto the $\mathbf{e_n}$-$\mathbf{e_{\perp}}$ plane. Points P and AP are the equilibrium positions of $\mathbf{M}$ along the easy axis. (c), (d) Schematics of MTJ energy density landscape (c) before and (d) during a voltage pulse, for $\mathbf{M}$ in plane ($\varphi=\pi/2$), where $K$ is the easy axis anisotropy energy density, $k_B T$ is the thermal energy at temperature $T$, and $v$ is the free layer volume. In (d) the field-like spin torque lowers the energy barrier allowing magnetization switching via a combination of Slonczewski's spin torque and random fluctuations.}\label{fig:fig1} \end{center}
\end{figure}

We consider a macrospin model corresponding to a rectangular MTJ
[Fig.~\ref{fig:fig1}(a)] with nominal free layer dimensions
100$\times$300$\times$2.5 nm$^3$, saturation magnetization $4\pi M_S$=4.4 kG, easy axis anisotropy field $2 K/M_S=H_K$=80 Oe, dipole
field coupling to the reference layer $H_{D}$= 28 Oe favoring the P
state, and with parallel (P) and antiparallel (AP) resistances
$R_P$=286 $\Omega$ and $R_{AP}$=364 $\Omega$ respectively, as quoted
in Ref.~\cite{devolder_PRL2008}. The magnetization orientation of
the free layer, given by the unit vector $\mathbf{M}$, is described by the angles $\theta$
and $\varphi$ defined in Fig.~\ref{fig:fig1}(b). In the absence of
external fields and voltage and ignoring any magnetic coupling
between the free and pinned layers, the equilibrium positions of
$\mathbf{M}$ lie on the easy axis. A schematic of the energy density
profile at $\varphi=\pi/2$, i.e. for in-plane $\mathbf{M}$, is
represented in Fig.~\ref{fig:fig1}(c). We generate a
Maxwell-Boltzmann distribution of 1000 initial orientations of
$\mathbf{M}$ corresponding to a temperature of 300~K. The time
evolution of the ensemble of trajectories is found by solving the
stochastic Landau-Lifshitz-Gilbert (LLG) equation in the presence of
both Slonczewski's~\cite{slonczewski_JMMM1996} and
field-like~\cite{li_PRL2008} spin torque terms,
\begin{equation}
T=a_J \mathbf{M}\times(\mathbf{M}\times\mathbf{m_P}) +b_J
\mathbf{M}\times\mathbf{m_P},
\end{equation}
for a macrospin nanomagnet with energy density
\begin{equation}
u = - K({\bf M}\cdot{\bf e}_{||})^2 - M_S(H_D+H_{\parallel}) {\bf
M}\cdot{\bf e}_{||}+K_P{\bf M}\cdot{\bf e}_{n}.
\end{equation}
The effects of thermal fluctuations during the evolution of
$\mathbf{M}$ are included by using thermal random
fields~\cite{brown_PHYSREV1963,garcia_PRB1998}. We describe the angle dependence of the
MTJ resistance by $R(\theta)=R_P+(R_{AP}-R_P)\sin^2\theta/2$.
Although the results shown here are for an angle independent
efficiency, $g(\theta,p)=\eta$, similar results were obtained by
including Slonczewski's MTJ efficiency, $g(\theta,p)=p/(2+2 p^2
\cos(\theta))$~\cite{slonczewski_PRB2005}, where $p$ is the spin
polarization of the tunneling electrons. As proposed by Li
\textit{et al.} \cite{li_PRL2008}, we use $b_J= \epsilon |V| a_J$, where V is the
voltage and $\epsilon$ controls the relative amplitude of the
two spin torque terms. Ref.~\cite{li_PRL2008} estimates $\epsilon\sim$1 V$^{-1}$ for typical magnetic materials. We model the voltage waveform as a step with
55 ps risetime.
$v$
By appropriately choosing the values of $\epsilon$, $\eta$, and the
damping, $\alpha$, it is possible to obtain a situation where
(i)~the field-like spin torque is small enough so that the P state
is still stable and a large majority of the thermally distributed
initial orientations of $\mathbf{M}$ are within the stability region
of P, as shown in Fig.~\ref{fig:fig1}(d), and (ii) the Slonczewski's
spin torque is too weak to induce magnetization reversal by itself.
Under these conditions, it is the combination of Slonczewski's
spin-torque together with thermal fluctuations which eventually push
the magnetization over the barrier, which has been lowered by the
field-like spin torque. From the values of $R_P$ and $R_{AP}$ we
obtain a zero bias tunneling magnetoresistance TMR$\approx$27\%,
from which the spin polarization $p\approx$34\% and efficiency
$\eta\approx$0.15 can be obtained~\cite{diao_APL2005}. As the
voltage across the MTJ is increased, the efficiency will decrease,
possibly by more than 70\% when V=1 V (the voltage used in
Ref.~\cite{devolder_PRL2008}).

\begin{figure}[tb]
\begin{center}
\includegraphics[width=3.5in]{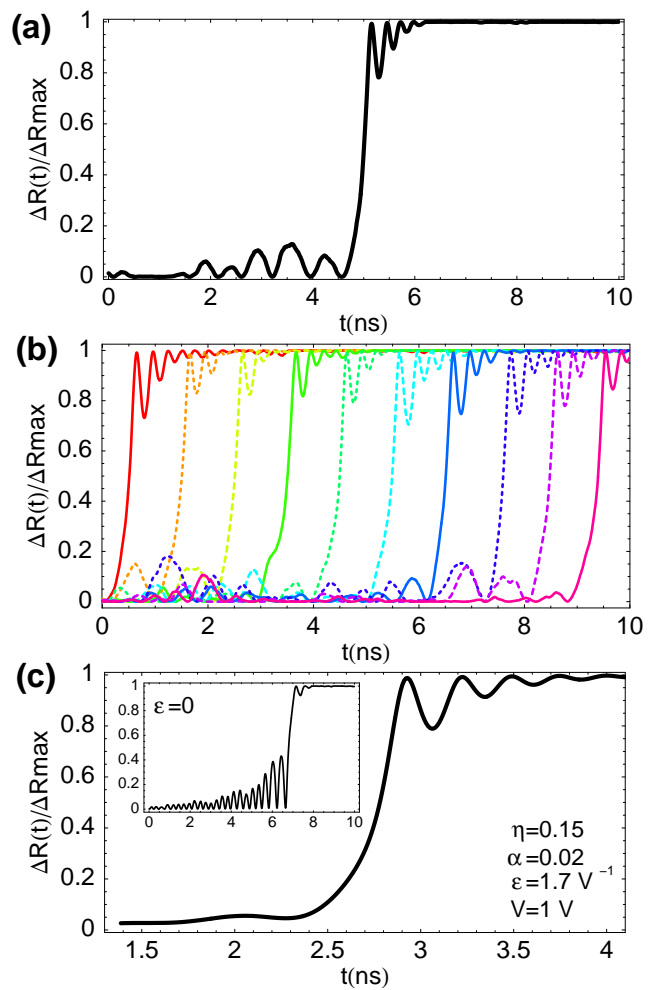}
\caption{(Color online) (a) Normalized change in resistance as a
function of time for a randomly chosen trajectory. (b) Same for a set of trajectories with switching times $t_S$
distributed between 0 and 10 ns. (c) Average of all resistance
traces with $t_S\geq\overline{t_S}$ after aligning their switching
time with the average value $\overline{t_S}$=2.75 ns. Inset:
switching via Slonczewski's spin torque only ($\epsilon=0$). All
other parameters are kept constant.}
 \label{fig:fig2}
 \end{center}
\end{figure}

We first find the time evolution of $\mathbf{M}$ for $\eta$=0.15,
$\alpha$=0.02, $\epsilon$=1.7 V$^{-1}$, and V=1 V. We generate 1000
trajectories such as the one shown in Fig.~\ref{fig:fig2}(a). Then we randomly pick ten of them, one by one, with the restriction that any picked trajectory has to have its switching
time clearly spaced from all previously picked trajectories. This process
ensures that the set is representative of the whole ensemble. The
time dependence of the resistance for the selected trajectories is
shown in Fig.~\ref{fig:fig2}(b) (compare with Fig. 4 in
Ref.~\cite{devolder_PRL2008}). We observe (i) a slow initial
increase in the resistance, (ii) random pre-switching fluctuations
and reproducible post-switching ringing, and (iii) similarity in the
behavior of $\mathbf{M}$ trajectories shifted so as to align their switching times.
Furthermore, even though in some traces the transition from a slow
resistance increase to a fast switch is subtle, clear transitions at
about 0.2$\Delta R_{max}$ are observable in many of the traces. We
note that this value corresponds to the resistance at the top of the
energy barrier (Fig.~\ref{fig:fig1}(d)) separating the P and AP
states. The average switching time for the ensemble of trajectories
is $\overline{t_S}$=2.8 ns, close to the value measured in
Ref.~\cite{devolder_PRL2008} at 1.1 V ($\overline{t_S}$=2.5 ns), but
smaller than what they observed at 1 V ($\overline{t_S}$=5.3 ns).
However, since the average switching time depends in part on the
function $g(\theta,p)$ and the value of $\epsilon$ used, which are
not well known, complete numerical agreement with the experimental
results is not expected. To confirm the randomness of the
pre-switching fluctuations and coherence of post-switching
oscillations we performed the following averaging. From the full
sample of 1000 trajectories we picked the ones with switching time
$t_S$ in the interval $\overline{t_S} \pm \delta t_S$ with $\delta
t_S$ being the standard deviation. We then shifted all resistance
traces to the same switching time $\overline{t_S}$ and averaged
them. The result (Fig.~\ref{fig:fig2}(c)) shows that the
post-switching ringing is indeed preserved but the random
fluctuations that precede magnetization reversal are averaged out.
It also shows that the transition from a slow to a sharp resistance
increase, which occurs close to 0.2$\Delta$R, is preserved. The
inset to Fig.~\ref{fig:fig2}(c) allows comparison of this result with
the typical switching process induced by Slonczewski's spin torque
alone ($\epsilon$=0). In the latter case one observes a clear build
up of the precession amplitude preceeding magnetization reversal,
while the post-switching oscillations are reduced.

\begin{figure}[tb]
\begin{center}
\includegraphics[width=3.5in]{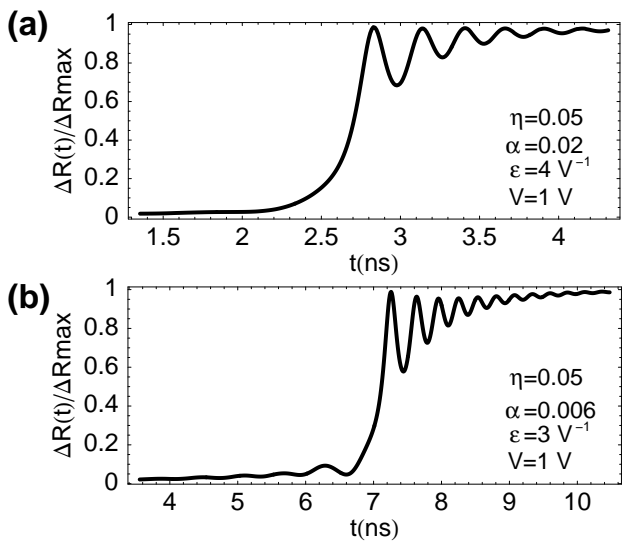}
\caption{(a), (b) Average of resistance traces after aligning their
switching times with $\overline{t_S}$. In (a) $\overline{t_S}$=2.7
ns, while in (b) $\overline{t_S}$=7.1 ns.}\label{fig:fig3}
\end{center}
\end{figure}

We tested the sensitivity of our observations to changes in the
parameters $\eta$, $\alpha$, and $\epsilon$. Figure \ref{fig:fig3}
shows averaged resistance traces for two sets of values of the
parameters $\eta$, $\alpha$, and $\epsilon$, illustrating that a
wide range of these parameters lead to the same general behavior. As
long as the efficiency is small enough so that Slonczewski's spin
torque is comparable to the random thermal torques (this condition
relates $\eta$ and $\alpha$), the behavior of the resistance is well
described by Fig.~\ref{fig:fig3}(a). Even in the case where P
becomes an energy maximum (for large values of $\epsilon$ and
$\eta$) we observe similar behavior, although the average switching
times are well below 1 ns, and the probability of having an initial
orientation of $\mathbf{M }$ with a switching time of more than 1 ns
is negligibly small. However, as long as $\epsilon$ is chosen so
that the P-AP energy barrier is a few times larger than the thermal
energy, it is possible to obtain a wide range of switching times,
even exceeding the value of 5.3 ns measured in
Ref.~\cite{devolder_PRL2008} [Fig.~\ref{fig:fig3}(b)]. We observe
that as the average switching times increase due to a larger P-AP
barrier, the amplitude of the random pre-switching fluctuations
increases, since larger fluctuations in $\theta$ (and therefore in
resistance) are required to overcome the barrier. As shown in
Fig.~\ref{fig:fig3}(b) some pre-switching oscillations still remain
after averaging, but their amplitude is much smaller than the
amplitude of the post-switching oscillations. Therefore, there
exists a large range of reasonable values for the parameters $\eta$,
$\alpha$, and $\epsilon$, which satisfy the conditions described
above and reproduce the main experimental observations.

\begin{figure}[tb]
\begin{center}
\includegraphics[width=3.5in]{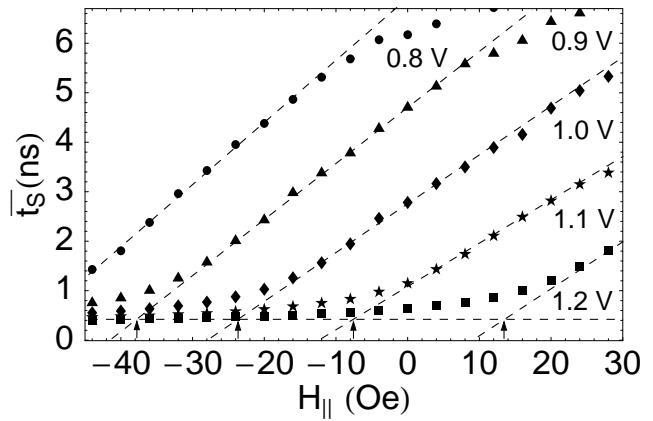}
\caption{Average switching time as a function of applied field for
voltages between 0.8 and 1.2 V in 0.1 V increments. The horizontal
dashed line shows the negative field saturation value, while the
tilted dashed lines are guides to the eye.}\label{fig:fig4}
\end{center}
\end{figure}

In the present device geometry an application of external magnetic
field $H_{||}$ along the magnetic easy axis effectively modifies the
field-like torque constant $b_J  \to b_J - H_{||}$ leaving the
$a_J$ unchanged. Experiments with separate control of the MTJ
voltage $V$ and applied field $H_{||}$ can provide valuable
information on the spin-transfer torque parameters. The behavior of
the average switching time predicted from simulations is shown in
Fig.~\ref{fig:fig4}. We used the same parameters as in
Fig.~\ref{fig:fig2}, but varied the voltage between 0.8 and 1.2 V
in steps of 0.1 V, and applied magnetic fields between -44 and 28
Oe along the easy axis. Positive fields favor the P state. The
apparent saturation of the switching time for the 0.8~V and 0.9~V traces
occurs since we mimic the averaging performed in an experiment with
a 10 ns current pulse (switching events with $t_S > 10$~ns are
excluded). To understand these simulations we note that both applied
field and voltage affect the energy barrier, with $b_J$ decreasing
and $H_{||}$ increasing it. Since $a_J = (\hbar \eta/2 e v M_s)V/R$
where $v$ is the free layer volume, and $R$ is the effective device
resistance, we get $b_J = \beta V^2$ with $\beta = 2 e R v
M_S/\hbar\eta\epsilon$. The barrier is completely eliminated at $H_K
+ H_D + H_{||} - b_J = 0$ which gives a crossover field
$$
H_{||*} = \beta V^2 - H_K - H_D,
$$
(marked by arrows in Fig.~\ref{fig:fig4}) separating the ultrafast and
normal switching regimes. Although the crossover is broad due to
finite temperature, one would be able to extract $H_{||*}$ from
experimental data by extrapolating the dependencies
$\overline{t_S}(H_{||})$ above and below the crossover. Then
plotting $H_{||*}$ as a function of $V^2$ one can find the
coefficient $\beta$ and extract the value of $\epsilon$, i.e., the
relative strength of the field-like spin torque term. To demonstrate the
feasibility of such data analysis we fitted the simulated data and
obtained $\epsilon \approx 1.75$ V$^{-1}$, $H_K + H_D \approx 107.5$~Oe for a
dataset generated with $\epsilon = 1.7$ V$^{-1}$ and $H_K + H_D = 108$~Oe.
This method should allow verification of the $V^2$ dependence of the
field-like torque term and the measurement of its magnitude,
characterized in our model by~$\epsilon$.

In conclusion, we show that by considering an additional field-like
spin torque of magnitude similar to Slonczewski's spin torque, a
macrospin description of MTJ switching reproduces the ``incubation
delay'', the lack of coherent magnetization precession before
reversal, and the post-switching oscillations observed in
Ref.~\cite{devolder_PRL2008}. Our analysis suggests an additional
experimental procedure to measure the voltage dependence of the
field-like spin torque term: an external magnetic field can be
applied along the easy axis to cancel or reinforce the field-like
spin torque, while the average switching time in response to voltage
pulses of different amplitudes is measured. In contrast to previous
studies based on careful analysis of the antisymmetric component of
ferromagnetic resonance spectra, or fits of the critical switching
voltage to the spin torque model, our proposed measurements can
provide a more direct access to the field-like spin torque term,
clarify its origin, and obtain its voltage dependence in a wider
range.

\bibliography{references}

\begin{thebibliography}{20}
\expandafter\ifx\csname natexlab\endcsname\relax\def\natexlab#1{#1}\fi
\expandafter\ifx\csname bibnamefont\endcsname\relax
  \def\bibnamefont#1{#1}\fi
\expandafter\ifx\csname bibfnamefont\endcsname\relax
  \def\bibfnamefont#1{#1}\fi
\expandafter\ifx\csname citenamefont\endcsname\relax
  \def\citenamefont#1{#1}\fi
\expandafter\ifx\csname url\endcsname\relax
  \def\url#1{\texttt{#1}}\fi
\expandafter\ifx\csname urlprefix\endcsname\relax\def\urlprefix{URL }\fi
\providecommand{\bibinfo}[2]{#2}
\providecommand{\eprint}[2][]{\url{#2}}

\bibitem[{\citenamefont{{Slonczewski}}(1996)}]{slonczewski_JMMM1996}
\bibinfo{author}{\bibfnamefont{J.~C.} \bibnamefont{{Slonczewski}}},
  \bibinfo{journal}{J. Magn. Magn. Mater.} \textbf{\bibinfo{volume}{159}},
  \bibinfo{pages}{L1} (\bibinfo{year}{1996}).

\bibitem[{\citenamefont{Berger}(1996)}]{berger_PRB1996}
\bibinfo{author}{\bibfnamefont{L.}~\bibnamefont{Berger}},
  \bibinfo{journal}{Phys. Rev. B} \textbf{\bibinfo{volume}{54}},
  \bibinfo{pages}{9353} (\bibinfo{year}{1996}).

\bibitem[{\citenamefont{Petit et~al.}(2007)\citenamefont{Petit, Baraduc,
  Thirion, Ebels, Liu, Li, Wang, and Dieny}}]{petit_PRL2007}
\bibinfo{author}{\bibfnamefont{S.}~\bibnamefont{Petit}},
  \bibinfo{author}{\bibfnamefont{C.}~\bibnamefont{Baraduc}},
  \bibinfo{author}{\bibfnamefont{C.}~\bibnamefont{Thirion}},
  \bibinfo{author}{\bibfnamefont{U.}~\bibnamefont{Ebels}},
  \bibinfo{author}{\bibfnamefont{Y.}~\bibnamefont{Liu}},
  \bibinfo{author}{\bibfnamefont{M.}~\bibnamefont{Li}},
  \bibinfo{author}{\bibfnamefont{P.}~\bibnamefont{Wang}}, \bibnamefont{and}
  \bibinfo{author}{\bibfnamefont{B.}~\bibnamefont{Dieny}},
  \bibinfo{journal}{Phys. Rev. Lett.} \textbf{\bibinfo{volume}{98}},
  \bibinfo{pages}{077203} (\bibinfo{year}{2007}).

\bibitem[{\citenamefont{{Kubota} et~al.}(2008)\citenamefont{{Kubota},
  {Fukushima}, {Yakushiji}, {Nagahama}, {Yuasa}, {Ando}, {Maehara}, {Nagamine},
  {Tsunekawa}, {Djayaprawira} et~al.}}]{kubota_NATPHYS2008}
\bibinfo{author}{\bibfnamefont{H.}~\bibnamefont{{Kubota}}},
  \bibinfo{author}{\bibfnamefont{A.}~\bibnamefont{{Fukushima}}},
  \bibinfo{author}{\bibfnamefont{K.}~\bibnamefont{{Yakushiji}}},
  \bibinfo{author}{\bibfnamefont{T.}~\bibnamefont{{Nagahama}}},
  \bibinfo{author}{\bibfnamefont{S.}~\bibnamefont{{Yuasa}}},
  \bibinfo{author}{\bibfnamefont{K.}~\bibnamefont{{Ando}}},
  \bibinfo{author}{\bibfnamefont{H.}~\bibnamefont{{Maehara}}},
  \bibinfo{author}{\bibfnamefont{Y.}~\bibnamefont{{Nagamine}}},
  \bibinfo{author}{\bibfnamefont{K.}~\bibnamefont{{Tsunekawa}}},
  \bibinfo{author}{\bibfnamefont{D.~D.} \bibnamefont{{Djayaprawira}}},
  \bibnamefont{et~al.}, \bibinfo{journal}{Nature Physics}
  \textbf{\bibinfo{volume}{4}}, \bibinfo{pages}{37} (\bibinfo{year}{2008}).

\bibitem[{\citenamefont{{Sankey} et~al.}(2008)\citenamefont{{Sankey}, {Cui},
  {Sun}, {Slonczewski}, {Buhrman}, and {Ralph}}}]{sankey_NATPHYS2008}
\bibinfo{author}{\bibfnamefont{J.~C.} \bibnamefont{{Sankey}}},
  \bibinfo{author}{\bibfnamefont{Y.-T.} \bibnamefont{{Cui}}},
  \bibinfo{author}{\bibfnamefont{J.~Z.} \bibnamefont{{Sun}}},
  \bibinfo{author}{\bibfnamefont{J.~C.} \bibnamefont{{Slonczewski}}},
  \bibinfo{author}{\bibfnamefont{R.~A.} \bibnamefont{{Buhrman}}},
  \bibnamefont{and} \bibinfo{author}{\bibfnamefont{D.~C.}
  \bibnamefont{{Ralph}}}, \bibinfo{journal}{Nature Physics}
  \textbf{\bibinfo{volume}{4}}, \bibinfo{pages}{67} (\bibinfo{year}{2008}).

\bibitem[{\citenamefont{Li et~al.}(2008)\citenamefont{Li, Zhang, Diao, Ding,
  Tang, Apalkov, Yang, Kawabata, and Huai}}]{li_PRL2008}
\bibinfo{author}{\bibfnamefont{Z.}~\bibnamefont{Li}},
  \bibinfo{author}{\bibfnamefont{S.}~\bibnamefont{Zhang}},
  \bibinfo{author}{\bibfnamefont{Z.}~\bibnamefont{Diao}},
  \bibinfo{author}{\bibfnamefont{Y.}~\bibnamefont{Ding}},
  \bibinfo{author}{\bibfnamefont{X.}~\bibnamefont{Tang}},
  \bibinfo{author}{\bibfnamefont{D.~M.} \bibnamefont{Apalkov}},
  \bibinfo{author}{\bibfnamefont{Z.}~\bibnamefont{Yang}},
  \bibinfo{author}{\bibfnamefont{K.}~\bibnamefont{Kawabata}}, \bibnamefont{and}
  \bibinfo{author}{\bibfnamefont{Y.}~\bibnamefont{Huai}},
  \bibinfo{journal}{Phys. Rev. Lett.} \textbf{\bibinfo{volume}{100}},
  \bibinfo{pages}{246602} (\bibinfo{year}{2008}).

\bibitem[{\citenamefont{Heide}(2001)}]{heide_PRL2001}
\bibinfo{author}{\bibfnamefont{C.}~\bibnamefont{Heide}},
  \bibinfo{journal}{Phys. Rev. Lett.} \textbf{\bibinfo{volume}{87}},
  \bibinfo{pages}{197201} (\bibinfo{year}{2001}).

\bibitem[{\citenamefont{Tserkovnyak et~al.}(2002)\citenamefont{Tserkovnyak,
  Brataas, and Bauer}}]{tserkovnyak_PRL2002}
\bibinfo{author}{\bibfnamefont{Y.}~\bibnamefont{Tserkovnyak}},
  \bibinfo{author}{\bibfnamefont{A.}~\bibnamefont{Brataas}}, \bibnamefont{and}
  \bibinfo{author}{\bibfnamefont{G.~E.~W.} \bibnamefont{Bauer}},
  \bibinfo{journal}{Phys. Rev. Lett.} \textbf{\bibinfo{volume}{88}},
  \bibinfo{pages}{117601} (\bibinfo{year}{2002}).

\bibitem[{\citenamefont{Tserkovnyak et~al.}(2005)\citenamefont{Tserkovnyak,
  Brataas, Bauer, and Halperin}}]{tserkovnyak_RMP2005}
\bibinfo{author}{\bibfnamefont{Y.}~\bibnamefont{Tserkovnyak}},
  \bibinfo{author}{\bibfnamefont{A.}~\bibnamefont{Brataas}},
  \bibinfo{author}{\bibfnamefont{G.~E.~W.} \bibnamefont{Bauer}},
  \bibnamefont{and} \bibinfo{author}{\bibfnamefont{B.~I.}
  \bibnamefont{Halperin}}, \bibinfo{journal}{Reviews of Modern Physics}
  \textbf{\bibinfo{volume}{77}}, \bibinfo{pages}{1375} (\bibinfo{year}{2005}).

\bibitem[{\citenamefont{Xia et~al.}(2002)\citenamefont{Xia, Kelly, Bauer,
  Brataas, and Turek}}]{xia_PRB2002}
\bibinfo{author}{\bibfnamefont{K.}~\bibnamefont{Xia}},
  \bibinfo{author}{\bibfnamefont{P.~J.} \bibnamefont{Kelly}},
  \bibinfo{author}{\bibfnamefont{G.~E.~W.} \bibnamefont{Bauer}},
  \bibinfo{author}{\bibfnamefont{A.}~\bibnamefont{Brataas}}, \bibnamefont{and}
  \bibinfo{author}{\bibfnamefont{I.}~\bibnamefont{Turek}},
  \bibinfo{journal}{Phys. Rev. B} \textbf{\bibinfo{volume}{65}},
  \bibinfo{pages}{220401} (\bibinfo{year}{2002}).

\bibitem[{\citenamefont{Zimmler et~al.}(2004)\citenamefont{Zimmler, \"Ozyilmaz,
  Chen, Kent, Sun, Rooks, and Koch}}]{zimmler_PRB2004}
\bibinfo{author}{\bibfnamefont{M.~A.} \bibnamefont{Zimmler}},
  \bibinfo{author}{\bibfnamefont{B.}~\bibnamefont{\"Ozyilmaz}},
  \bibinfo{author}{\bibfnamefont{W.}~\bibnamefont{Chen}},
  \bibinfo{author}{\bibfnamefont{A.~D.} \bibnamefont{Kent}},
  \bibinfo{author}{\bibfnamefont{J.~Z.} \bibnamefont{Sun}},
  \bibinfo{author}{\bibfnamefont{M.~J.} \bibnamefont{Rooks}}, \bibnamefont{and}
  \bibinfo{author}{\bibfnamefont{R.~H.} \bibnamefont{Koch}},
  \bibinfo{journal}{Phys. Rev. B} \textbf{\bibinfo{volume}{70}},
  \bibinfo{pages}{184438} (\bibinfo{year}{2004}).

\bibitem[{\citenamefont{Theodonis et~al.}(2006)\citenamefont{Theodonis,
  Kioussis, Kalitsov, Chshiev, and Butler}}]{theodonis_PRL2006}
\bibinfo{author}{\bibfnamefont{I.}~\bibnamefont{Theodonis}},
  \bibinfo{author}{\bibfnamefont{N.}~\bibnamefont{Kioussis}},
  \bibinfo{author}{\bibfnamefont{A.}~\bibnamefont{Kalitsov}},
  \bibinfo{author}{\bibfnamefont{M.}~\bibnamefont{Chshiev}}, \bibnamefont{and}
  \bibinfo{author}{\bibfnamefont{W.~H.} \bibnamefont{Butler}},
  \bibinfo{journal}{Phys. Rev. Lett.} \textbf{\bibinfo{volume}{97}},
  \bibinfo{pages}{237205} (\bibinfo{year}{2006}).

\bibitem[{\citenamefont{{Slonczewski} and {Sun}}(2007)}]{slonczewski_JMMM2007}
\bibinfo{author}{\bibfnamefont{J.~C.} \bibnamefont{{Slonczewski}}}
  \bibnamefont{and} \bibinfo{author}{\bibfnamefont{J.~Z.} \bibnamefont{{Sun}}},
  \bibinfo{journal}{J. Magn. Magn. Mater.} \textbf{\bibinfo{volume}{310}},
  \bibinfo{pages}{169} (\bibinfo{year}{2007}).

\bibitem[{\citenamefont{Devolder et~al.}(2008)\citenamefont{Devolder, Hayakawa,
  Ito, Takahashi, Ikeda, Crozat, Zerounian, Kim, Chappert, and
  Ohno}}]{devolder_PRL2008}
\bibinfo{author}{\bibfnamefont{T.}~\bibnamefont{Devolder}},
  \bibinfo{author}{\bibfnamefont{J.}~\bibnamefont{Hayakawa}},
  \bibinfo{author}{\bibfnamefont{K.}~\bibnamefont{Ito}},
  \bibinfo{author}{\bibfnamefont{H.}~\bibnamefont{Takahashi}},
  \bibinfo{author}{\bibfnamefont{S.}~\bibnamefont{Ikeda}},
  \bibinfo{author}{\bibfnamefont{P.}~\bibnamefont{Crozat}},
  \bibinfo{author}{\bibfnamefont{N.}~\bibnamefont{Zerounian}},
  \bibinfo{author}{\bibfnamefont{J.-V.} \bibnamefont{Kim}},
  \bibinfo{author}{\bibfnamefont{C.}~\bibnamefont{Chappert}}, \bibnamefont{and}
  \bibinfo{author}{\bibfnamefont{H.}~\bibnamefont{Ohno}},
  \bibinfo{journal}{Phys. Rev. Lett.} \textbf{\bibinfo{volume}{100}},
  \bibinfo{pages}{057206} (\bibinfo{year}{2008}).

\bibitem[{\citenamefont{Sun}(2000)}]{sun_PRB2000}
\bibinfo{author}{\bibfnamefont{J.~Z.} \bibnamefont{Sun}},
  \bibinfo{journal}{Phys. Rev. B} \textbf{\bibinfo{volume}{62}},
  \bibinfo{pages}{570} (\bibinfo{year}{2000}).

\bibitem[{\citenamefont{Krivorotov et~al.}(2005)\citenamefont{Krivorotov,
  Emley, Sankey, Kiselev, Ralph, and Buhrman}}]{krivorotov_SCIENCE2005}
\bibinfo{author}{\bibfnamefont{I.~N.} \bibnamefont{Krivorotov}},
  \bibinfo{author}{\bibfnamefont{N.~C.} \bibnamefont{Emley}},
  \bibinfo{author}{\bibfnamefont{J.~C.} \bibnamefont{Sankey}},
  \bibinfo{author}{\bibfnamefont{S.~I.} \bibnamefont{Kiselev}},
  \bibinfo{author}{\bibfnamefont{D.~C.} \bibnamefont{Ralph}}, \bibnamefont{and}
  \bibinfo{author}{\bibfnamefont{R.~A.} \bibnamefont{Buhrman}},
  \bibinfo{journal}{Science} \textbf{\bibinfo{volume}{307}},
  \bibinfo{pages}{228} (\bibinfo{year}{2005}).

\bibitem[{\citenamefont{Brown}(1963)}]{brown_PHYSREV1963}
\bibinfo{author}{\bibfnamefont{W.~F.} \bibnamefont{Brown}},
  \bibinfo{journal}{Phys. Rev.} \textbf{\bibinfo{volume}{130}},
  \bibinfo{pages}{1677} (\bibinfo{year}{1963}).

\bibitem[{\citenamefont{Garc\'ia-Palacios and L\'azaro}(1998)}]{garcia_PRB1998}
\bibinfo{author}{\bibfnamefont{J.~L.} \bibnamefont{Garc\'ia-Palacios}}
  \bibnamefont{and} \bibinfo{author}{\bibfnamefont{F.~J.}
  \bibnamefont{L\'azaro}}, \bibinfo{journal}{Phys. Rev. B}
  \textbf{\bibinfo{volume}{58}}, \bibinfo{pages}{14937} (\bibinfo{year}{1998}).

\bibitem[{\citenamefont{Slonczewski}(2005)}]{slonczewski_PRB2005}
\bibinfo{author}{\bibfnamefont{J.~C.} \bibnamefont{Slonczewski}},
  \bibinfo{journal}{Phys. Rev. B} \textbf{\bibinfo{volume}{71}},
  \bibinfo{pages}{024411} (\bibinfo{year}{2005}).

\bibitem[{\citenamefont{Diao et~al.}(2005)\citenamefont{Diao, Apalkov, Pakala,
  Ding, Panchula, and Huai}}]{diao_APL2005}
\bibinfo{author}{\bibfnamefont{Z.}~\bibnamefont{Diao}},
  \bibinfo{author}{\bibfnamefont{D.}~\bibnamefont{Apalkov}},
  \bibinfo{author}{\bibfnamefont{M.}~\bibnamefont{Pakala}},
  \bibinfo{author}{\bibfnamefont{Y.}~\bibnamefont{Ding}},
  \bibinfo{author}{\bibfnamefont{A.}~\bibnamefont{Panchula}}, \bibnamefont{and}
  \bibinfo{author}{\bibfnamefont{Y.}~\bibnamefont{Huai}},
  \bibinfo{journal}{Appl. Phys. Lett.} \textbf{\bibinfo{volume}{87}},
  \bibinfo{pages}{232502} (\bibinfo{year}{2005}).

\end{thebibliography}
\end{document}